# Design and analysis of a 1-ton prototype of the Jinping Neutrino Experiment

Zongyi Wang[a,*], Yuanqing Wang[b], Zhe Wang[c], Shaomin Chen[c], Xinxi Du[a]
Tianxiong Zhang[d], Ziyi Guo[c] and Huanxin Yuan[a]

[a] School of Civil Engineering, Wuhan University, Wuhan 430072, China;
[b] Key Laboratory of Civil Engineering Safety and Durability of Education Ministry, Tsinghua University, Beijing 100084, China;
[c] Department of Engineering Physics, Tsinghua University, Beijing 100084, China;
[d] School of Civil Engineering, Tianjin University, Tianjin 300072, China;

*Corresponding author: wangzongyi1990@outlook.com

**Abstract:** The Jinping Neutrino Experiment will perform an in-depth research on solar neutrinos and geo-neutrinos. Two structural options (i.e., cylindrical and spherical schemes) are proposed for the Jinping detector based on other successful underground neutrino detectors. Several key factors in the design are also discussed in detail. A 1-ton prototype of the Jinping experiment is proposed based on physics requirements. Subsequently, the structural design, installation procedure, and mechanical analysis of the neutrino detector prototype are discussed. The results show that the maximum Mises stresses on the acrylic vessel, stainless steel truss, and the tank are all lower than the design values of the strengths. The stability requirement of the stainless steel truss in the detector prototype is satisfied. Consequently, the structural scheme for the 1-ton prototype is safe and reliable.

## 1. Introduction

The Jinping Neutrino Experiment (hereafter referred to as the Jinping experiment) was proposed in 2015 [1], and several related studies were done thereafter [2-5]. With the advantage of the China Jinping Underground Laboratory (CJPL), which is located 2400 m under the Jinping Mountain in Sichuan, China, the Jinping experiment will perform an in-depth research on the solar neutrino and geo-neutrino physics. The experiment is expected to measure the fluxes of several solar neutrino components, including the carbon-nitrogen-oxygen (CNO) cycle neutrinos, further study the neutrino oscillation in the sun's high-density environment, and address the solar metallicity problem. Jinping is also expected to precisely measure the flux of geo-neutrinos and the Th/U ratio in the Earth and help disentangle the U and Th distribution in the crust and mantle.

The fiducial mass of the Jinping detector is expected to reach 2000 tons for the solar neutrino physics study [1], which is equivalently 3000 tons for geo-neutrino physics. For the double-detector scheme, each detector has a scale of over 20 m in dimension, which raises a challenge in detector design. The requirement of a radioactively pure detector also causes trouble in selecting the proper materials and structures.

A 1-ton prototype is designed to test the performance of several related key detector components, to understand the neutrino detection technology, and to measure the underground background level.

Accordingly, Section 2 introduces the proposed two options for the Jinping detector structure obtained after learning from other successful underground neutrino detectors. Section 3 discusses the physical requirements. Section 4.1 describes the structure of the physical 1-ton prototype, while Section 4.2 introduces the installation process. The acrylic vessel, stainless steel truss, and stainless steel tank are separately modeled and analyzed in the finite element analysis (FEA) software package, ABAQUS, in Section 4.3. Section 5 presents the conclusions.

## 2. Consideration of the Jinping detector design

The collaboration plans to build two detectors in the experimental hall of CJPL (Fig. 1) with liquid scintillator or slow liquid scintillator [2-5] as a neutrino target and detection material. Each detector for the solar neutrino study has a fiducial target mass of 1 kt, which is equivalent to 1.5 kt for geo-neutrinos and supernova neutrinos.

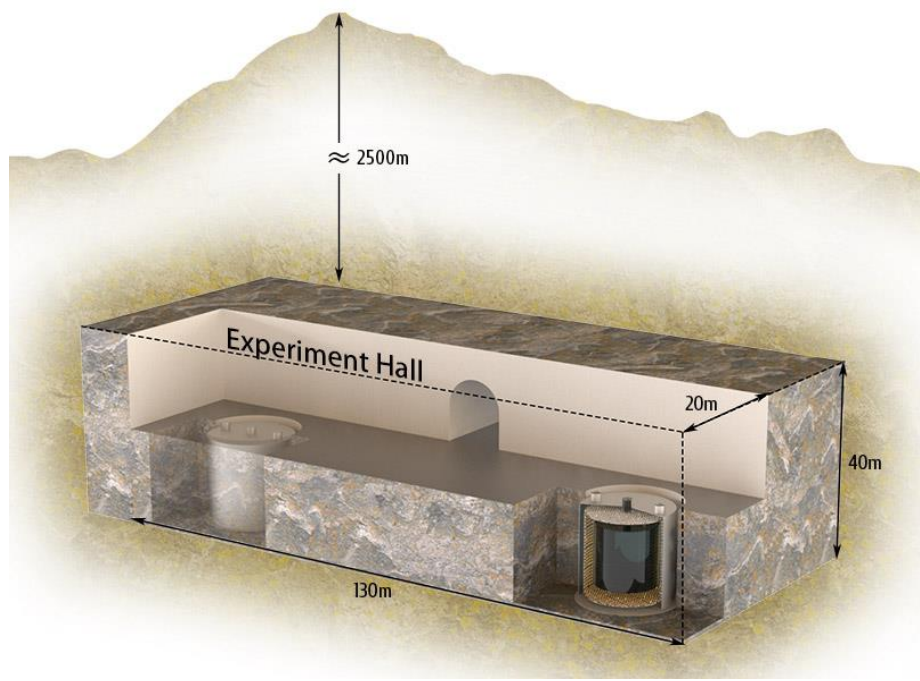

Fig. 1 Schematic diagram of the experimental hall

The underground neutrino detectors built worldwide generally adopt cylindrical (e.g, Super Kamiokande [6], Daya Bay [7], RENO [8], Double Chooz [9], and NEMO [10]) and spherical (e.g., SNO [11], JUNO [12-13], Borexino [14], and Kamland [15]) forms.

As a representative of the cylindrical structure, we discuss the Daya Bay detector [7]. It is split into three layers as follows: inner acrylic, outer acrylic, and stainless steel tank. The inner acrylic contains 20 tons of liquid scintillator and 192 photomultipliers (PMTs) are mounted on the inner side of the stainless steel tank. The SNO detector [11] is an example of spherical detector comprising an acrylic sphere that is 12 m in diameter and a stainless steel truss that is 17.8 m in diameter. The acrylic sphere has the capacity to contain 1 kt of heavy water and is supported through 10 rope loops made of synthetic fiber. In addition, 9438 inward-facing PMTs are mounted on the stainless steel truss.

The cylindrical and spherical schemes for the Jinping detector are adopted from these successful detectors (Fig. 2). In the cylindrical scheme, both the diameter and height of the acrylic vessel are 14 m. The vessel is supported by an outer single-layer stainless steel truss, and both the truss diameter and height are 20 m. Ropes made of synthetic fiber connect the acrylic vessel and the truss. The PMTs are uniformly distributed on the stainless steel truss. They detect the light produced in the acrylic vessel.

For the spherical scheme, the sphere has a diameter of 14 m, while the stainless steel truss has 20 m diameter. Compared to the cylindrical scheme, the spherical scheme is more feasible for the structural design, because of the uniform stress distribution on the spherical vessel. But the cylindrical scheme has an advantage of making full use of the space and can contain more liquid scintillator.

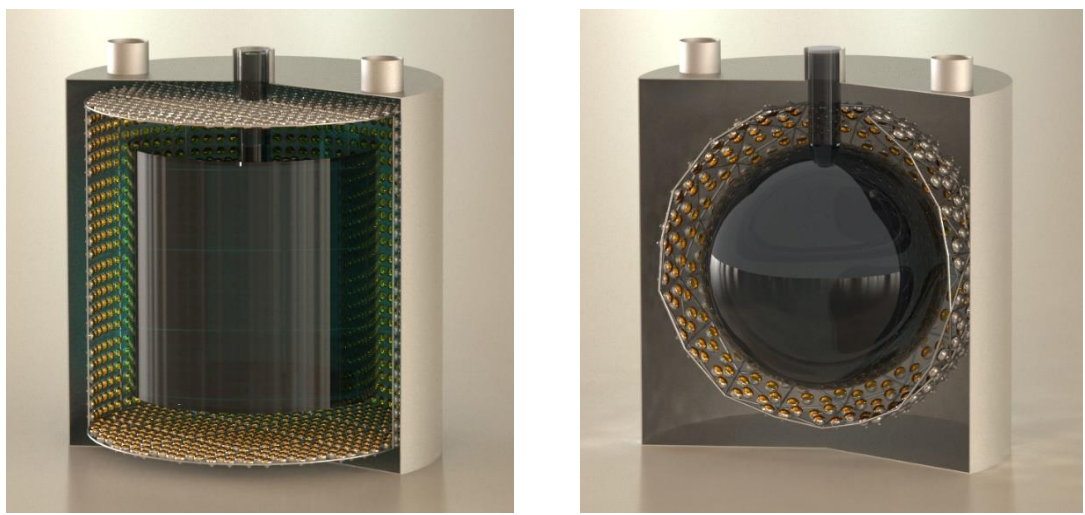

(a) Cylindrical scheme  (b) Spherical scheme

Fig. 2 Schematic diagram of the Jinping detector

In the operational period, the liquid scintillator (typical density: $0.866 \times 10^3$ kg/m$^3$) is inside the acrylic vessel, and purified water is outside (density: $10^3$ kg/m$^3$). Accordingly, three following working conditions need to be considered when designing the structural scheme: (a) self-weight of the structure before the filling; (b) filling process; and (c) condition after the filling. The allowable stress level, stability problem, and connection between acrylic and fiber ropes are vital factors that we must consider in the structural design. In view of the 30-year service life, various influential factors (e.g., earthquake, liquid level vibration and temperature change) must be considered to ensure the safety of the Jinping detector.

The following are some of the basic inputs for consideration:

1) Acrylic plastic is a kind of polymer material. Its response to external force

varies with temperature, stress level, and so on. The tensile strength of the acrylic material at the standard test condition is approximately 50─90 MPa [16-18]. When subjected to a sustained load for a long time, the acrylic will probably fail even at a low stress level because of creep. Stachiw [16] proposed a conservative allowable working stress level (i.e., 10.335 MPa) for the acrylic material. An additional safety margin should be respected considering the importance of Jinping detector. A creep test for the acrylic material used in the Jinping detector will be conducted in the future. Moreover, the allowable stress level will be determined. As a reference for the calculation, we simply suppose that the safety of the acrylic vessel can be ensured by limiting the Mises stress to 5 MPa.

2) Stainless steel has a higher strength compared to the acrylic material, and no pronounced creep. As a supporting structure, the stainless steel truss must satisfy the mechanical requirement. It is planned to use S31608 stainless steel in the Jinping detector. The corresponding yield strength is equal to 205 MPa, while the design value of the strength is 175 MPa [19]. The maximum stress on the stainless steel truss should be limited to the design value.

3) The stability problem of the stainless steel truss is greatly important because it is directly related to the structure's safety. According to the Chinese code "Technical specification for space frame structures" [20], the safety factor of the single layer truss is 4.2 if the elasticity is merely considered in the stability analysis. However, the safety factor is 2.0 if we perform an elastoplastic calculation.

4) The acrylic vessel is supported through fiber ropes. Their connections greatly affect the detector safety. The stress distribution on the acrylic vessel should be uniform, and stress concentration must be avoided. The strength of the fiber ropes under the long-term load also attracts our attention. A creep failure of ropes during operation is unacceptable. Accordingly, tests on the bearing capacity of the connection between the acrylic and the fiber rope will be performed in the future. The mechanical behavior of the synthetic fiber will be intensively studied.

5) According to the code for the seismic design of buildings [21], the seismic precautionary intensity of Xichang City, where the Jinping detector is located, should be at least 9 degrees. In addition, the design basic acceleration of ground motion should be greater than 0.40g. A structural calculation under seismic action is necessary.

6) In the filling process, the heights of the liquid levels inside and outside the acrylic vessel may change, which will lead to a variation of the stress distribution on the acrylic. Therefore, the liquid level variation is another influence factor that should be considered.

7) A temperature change will trigger an additional kind of stress, that is, temperature stress. Based on the test data, the increase and decrease of the temperature in the working environment of the Jinping detector will not exceed 10 ℃. Therefore, the evaluation of a temperature variation of 10 ℃ is relatively conservative.

## 3. Physical requirement of the 1-ton prototype of the Jinping detector

A 1-ton prototype is designed to test the performance of several related key detector components, to understand the neutrino detection technology with liquid scintillator and slow liquid scintillator, and to measure the underground background level.

The light yield and timing properties of the liquid scintillator and the slow liquid scintillator must be measured. The light yield is related to the energy resolution of the neutrino signal detection. The emission time of the scintillation light of the slow liquid scintillator is the key to particle identification.

The fast neutron background level can be measured with the 1-ton detector. The environmental gamma background can also be measured.

Aside from the physics properties of the detector, we are also interested in the design, construction and operation of such a detector, including the following: 1) handling of the liquid scintillator or the slow liquid scintillator; 2) optical and mechanical properties of the acrylic vessel; 3) light shielding; 4) photon detector (PMT) performance; 5) design and construction of the stainless steel supporting structure; 6) design and construction of stainless vessel; 7) pure water operation.

The scale of the 1-ton detector is necessary in realizing the abovementioned purpose, matching the required thickness to stop fast neutrons and fully contain gamma backgrounds.

## 4. Structural scheme of the 1-ton prototype of the Jinping detector

### 4.1 Structural design

A structural scheme of the combination of the stainless steel tank, stainless steel truss, and acrylic vessel is designed for the 1-tone prototype of the Jinping detector, based on physics requirements (Fig. 3). The tank has a diameter of 2000 mm, a thickness of 4 mm, and a height of 2090 mm. The height of the operational space in the working room is less than 4 m. Hence, the processing of the tank as a whole part will bring some problems in construction. Consequently, we divide the tank into three parts from top to bottom with corresponding heights equal to 470 mm, 800 mm and 820 mm, respectively. The adjacent parts are connected by flange plates. The stiffening ribs are uniformly distributed on the tank surface. Five filling holes, which are 114 mm in height and 242 in diameter, can be found on top of the tank. The middle filling hole is set for the acrylic vessel. Two holes are reserved for the cables of the PMTs, while the other two are used to fill the tank.

The function of the stainless steel truss is to support the acrylic vessel and fix the PMTs. The truss, which is 1795 mm in diameter and 1820 mm in height, is split into ten sections along the perimeter and three parts from top to bottom. The corresponding heights of the three parts are 540 mm, 740 mm, and 540 mm, respectively. A total of 10 diagonal braces are added on the top and bottom parts to increase the stability and bearing capacity of the structure. All the components adopt the solid steel bar with a diameter of 26 mm. Despite the poor stability, steel bars have the advantages of short processing cycle and high yield strength compared with steel tubes.

The acrylic vessel is divided into three parts as follows: sphere, chimney, and overflow tank. These parts are connected with each other through bulk polymerization. The chimney is designed for filling, while the overflow tank is used to prevent a spillover of the liquid caused by the temperature variation. The inner diameter of the sphere is equal to 1290 mm and the thickness is 20 mm. The inner diameter of the chimney is equal to 80 mm and the thickness is 35 mm. In the normal working condition, the liquids inside and outside the acrylic vessel are purified water. The inner level reaches one-third the height of the overflow tank. The acrylic vessel is principally supported by the stainless steel truss on the upper end, while an acrylic bearing on the bottom serves as an ancillary support. A soft pipe is connected with the bottom of the acrylic vessel to ensure the water circulation inside.

Three ropes fix the acrylic sphere in the horizontal direction (Fig. 4). This constraint is necessary because the acrylic vessel could vibrate in the transportation process, which is highly dangerous. Six acrylic slots are uniformly distributed on the sphere's equator. Each rope twines around the sphere through slots. Both ends are fixed on the stainless steel truss.

A sphere made of black acrylic material is located as a shield between the acrylic vessel and stainless steel truss. The shielding sphere is 10 mm thick. It lowers radiation from stainless steel materials in the inner tank. The shielding sphere has one hole for each PMT that detects the light produced in the vessel. The two poles of the shielding sphere are fixed. The upper end is connected with the vessel chimney, whereas the lower end is connected with the acrylic bearing. In addition, the shielding sphere's restriction in the horizontal direction is similar to that of the acrylic vessel.

A total of 30 PMTs are utilized in the 1-ton prototype of the Jinping detector. Half of the PMTs are Hamamatsu R5912, and the other half are HZC XP1805. The PMTs are alternately installed in four storeys. Five PMTs can be found in the top and bottom storeys, and 10 PMTs in the middle storeys. All the PMTs are fixed on the truss through stainless steel brackets. The uniform distribution of the PMTs is essential for the light measurement. In Fig. 5, letters A to E represent the PMTs, and letter O represents the center of the sphere. The COE and COD angles are 34.4°. The AOC angle is 40°, while that of the AOB is 39.8°.

**4.2 Installation process**

Fig. 6 shows the detector installation process. We first fix the acrylic vessel in process (a), then mount the shielding sphere in process (b). Subsequently, we install the acrylic vessel and the shielding sphere onto the stainless steel truss in process (c) and the PMTs in process (d). We hoist the whole inner structure into the bottom part of the tank in process (e) and finally connect the middle and top parts of the tank with the bottom part in process (f).

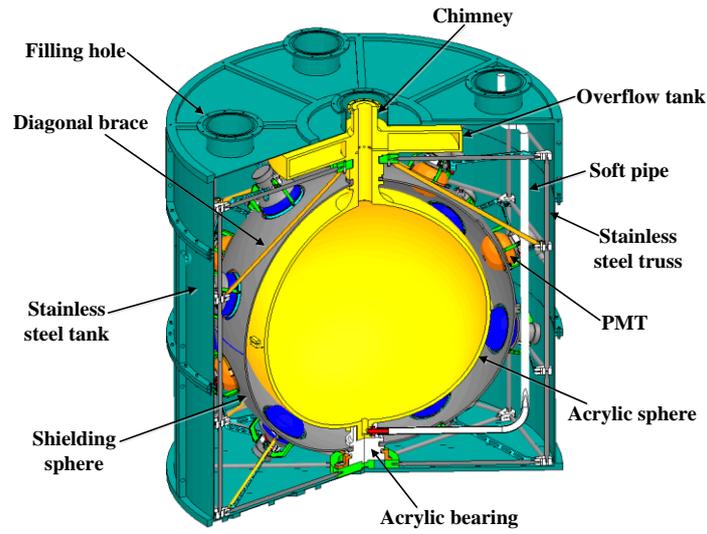

Fig. 3 1-Ton prototype of the Jinping detector

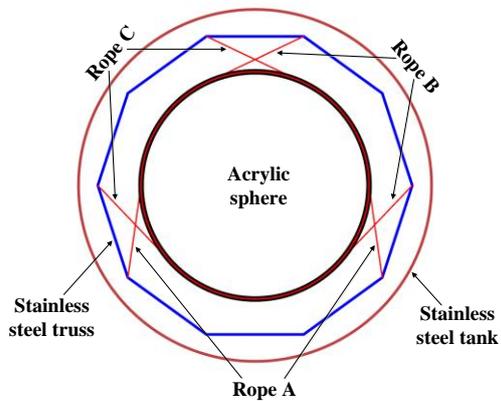

Fig. 4 Horizontal restriction of the acrylic sphere

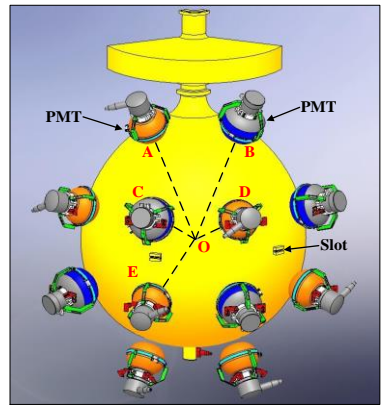

Fig. 5 Distribution of the PMTs

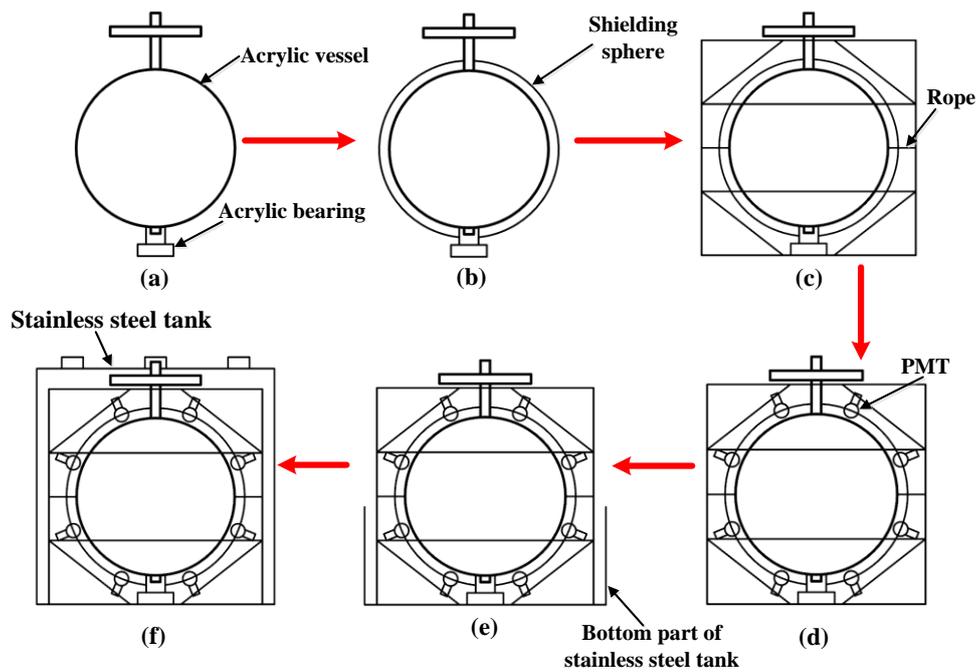

Fig. 6 Detector installation process

## 4.3 Structural calculation
### 4.3.1 Acrylic vessel

As previously mentioned, the maximum Mises stress on the acrylic should be limited to 5 MPa considering the creep of the acrylic material. Two working conditions are considered in the design of the acrylic vessel: (a) the stainless steel tank and the acrylic vessel are filled with purified water in the normal working condition (Working Condition 1); and (b) the acrylic vessel is filled with purified water and the water outside is lost due to a leak in the stainless steel tank (Working Condition 2). It is worth mentioning that the overflow tank is only utilized to catch the liquid spillover caused by temperature variation and has little influence on the bearing capacity of the acrylic vessel. Therefore, we ignore the overflow tank in the FEA for simplicity. The acrylic vessel is supported by the stainless steel truss on the top. The acrylic bearing under the vessel serves as an ancillary support. Therefore, the acrylic bearing on the bottom is ignored to consider the most dangerous condition.

The acrylic vessel is modeled in the FEA software, ABAQUS. The geometrical dimensions are determined according to the current design. The solid element C3D8 is used in the simulation. The three following kinds of loads are applied on the vessel under Working Condition 1: inner pressure, outer pressure, and self-weight. Under Working Condition 2, only two kinds of loads are considered: inner pressure and self-weight. In the FEA, the acrylic material density is equal to $1.2 \times 10^3$ kg/m$^3$, the modulus of elasticity is set to 2.77 GPa; and the Poisson's ratio is 0.376. Fig. 7 shows the results of calculation.

The figure shows that the maximum Mises stress under Working Condition 1 is 0.04 MPa, which is very low. The maximum Mises stress under Working Condition 2 is approximately 2.1 MPa. The value is significantly higher than that under Condition 1, while it is still lower than 5 MPa. Therefore, the acrylic vessel is safe to operate.

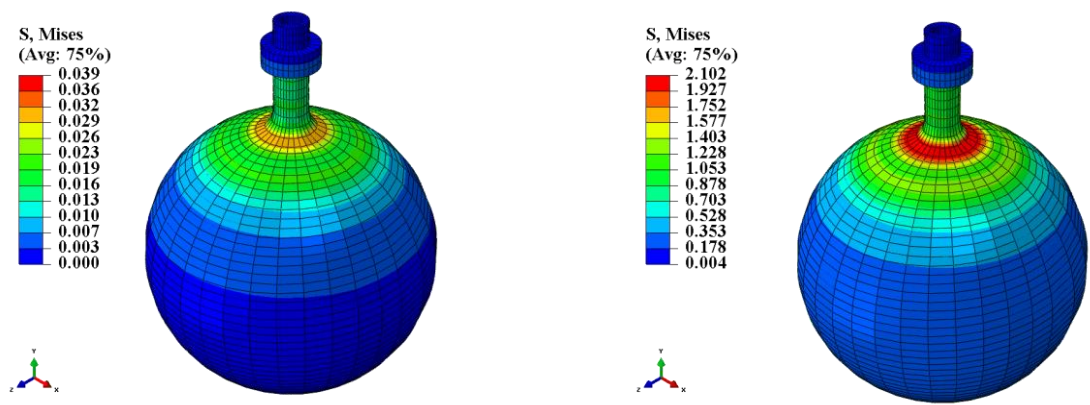

(a) Working Condition 1: the tank and acrylic vessel are filled with purified water

(b) Working Condition 2: the acrylic vessel is filled with purified water and no water outside the vessel

Fig. 7 Stress distribution on the acrylic vessel (MPa)

### 4.3.2 Stainless steel truss

The aforementioned analysis shows that Working Condition 2 is the worst condition. Accordingly, Working Condition 2 is considered when conducting the mechanical calculation of the stainless steel truss. All the components are stainless steel bars measuring 26 mm in diameter. The maximum length of the bars is 1059 mm, which corresponds to a slenderness ratio of 141.2. The value is smaller than 150 and satisfies the requirement of the Chinese code "Technical specification for space frame structures" [20]. The beam element B31 is utilized for the bar simulation. The constitutive relationship of the stainless steel adopts the elastic perfectly plastic model. According to the code "Technical specification for stainless steel structures", the yield strength is set as 205 MPa; the modulus of elasticity is 193 GPa; and Poisson's ratio is 0.3 [19].

The loads applied on the stainless steel truss include the following: (a) self-weight of the truss, (b) weights of the PMTs, and (c) pressure from the acrylic vessel filled with purified water. The weight of each PMT is 3 kg. It is applied at the truss joint as a point load. The pressure from the vessel is applied on the truss neck.

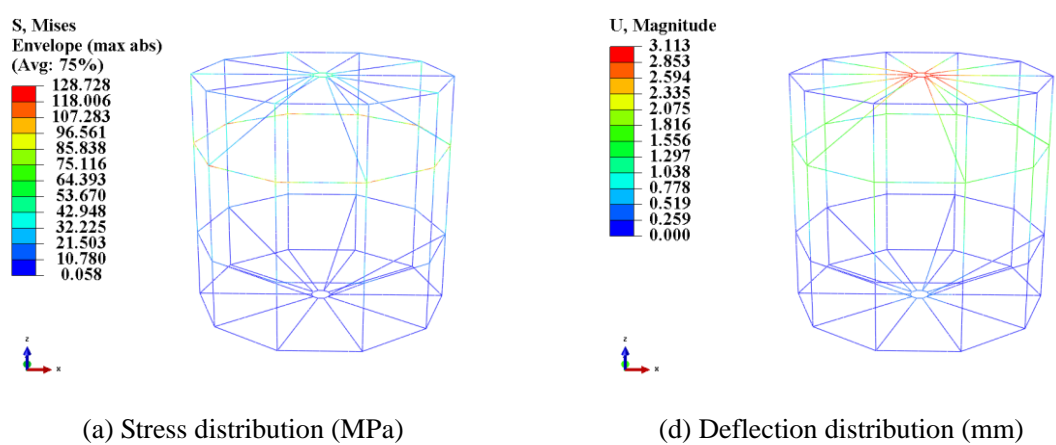

(a) Stress distribution (MPa)　　　　　　　(d) Deflection distribution (mm)

Fig. 8 Calculating results of the stainless steel truss

Fig.8 shows the results of calculation. The maximum Mises stress on the truss is 128.7 MPa, which is lower than the design goal of 175 MPa. The maximum deflection is 3.1 mm, which is approximately 1/635 of the truss diameter and smaller than 1/400. Therefore, these results satisfy the requirement of the Chinese code [20].

The stability problem of the stainless steel truss is of great importance. An elastic Buckle analysis is conducted. The first bucking mode is shown in Fig. 9(a). The first modal coefficient is equal to 23.2, which is significantly larger than 4.2. Subsequently, the calculation is performed using the Riks analysis method considering the geometrical and material nonlinearities. The initial imperfection distribution follows the first buckling mode. The maximum imperfection is set to 1/300 of the truss span (diameter). Fig. 9(b) depicts the load factor versus deflection curve of some feature point on the truss. It is indicated that the instability form of the truss is the snap-through buckling. The first buckling appears when the load factor reaches 6.9. Thereafter, the bearing capacity of the truss increases with the load factor. The second buckling appears when the load factor is increased to 8.4. The safety factor is equal to 6.9 and significantly larger than the required value of 2.0 in the code [20]. Therefore, the stainless steel truss is safe and reliable.

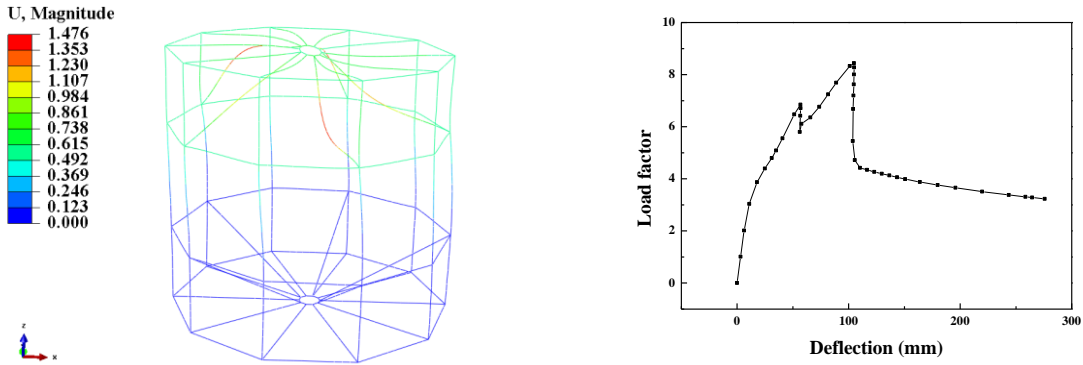

(a) First buckling mode  (b) Load factor versus deflection curve of the feature point

Fig. 9 Stability calculation of the stainless steel truss

### 4.3.3 Stainless steel tank

In the operational condition, the stainless steel tank is filled with purified water, and is, therefore, subjected to the lateral pressure from the water. We simply regard the stainless steel tank as a whole part and ignore the stiffing ribs in the FE model, which is conservative for the calculation.

The tank, which is 4 mm in thickness, is simulated by the shell element, S4R. The material properties are the same as those used in Section 4.3.2. For the load, we can simulate the water pressure by defining a linear pressure function. In addition, the constraint is imposed on the tank bottom in the calculation.

Fig. 10 shows the result. The maximum Mises stress on the tank is equal to 5.5 MPa. That is significantly lower than the design goal of 175 MPa. Consequently, the stainless steel tank is safe to operate.

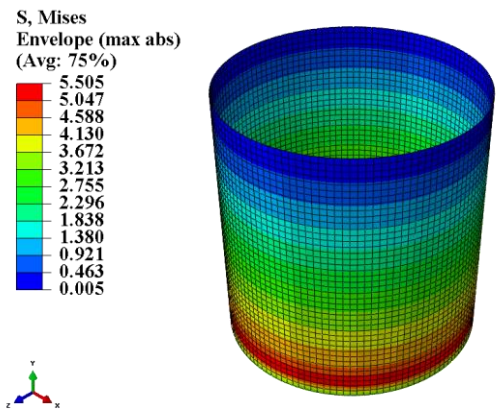

Fig. 10 Stress distribution on the stainless steel tank (MPa)

## 5. Conclusion

The Jinping experiment will make full use of the environmental condition and has advantages to study the solar neutrinos and geo-neutrinos. Two structural options for the Jinping detector are proposed based on the design of other successful underground neutrino detectors. The structural scheme of the physical 1-ton prototype of the detector is then designed and modeled in the present paper.

The cylindrical and spherical schemes are the two options for the Jinping detector. Compared with the spherical scheme, the cylindrical scheme has the advantage of making full use of the underground space. It is thus able to contain more detecting liquid. But, the spherical scheme is more feasible for the structural point-of-view, because of its uniform stress distribution. Both schemes use the combination of acrylic, fiber, and stainless steel. The maximum Mises stress on the acrylic should be limited to a low level in operation because of the creep. We plan to use S31608 stainless steel. The corresponding design value of the strength is 175 MPa. The Jinping detector is a kind of spatial structure. Hence, the stability problem needs to be considered. The safety factor should be larger than 4.2 in the elastic analysis and should be higher than 2.0 in the elastoplastic analysis. The influences of an earthquake, liquid level variation, and temperature change on the structure also need to be considered in the design.

The maximum Mises stress on the acrylic for the physical 1-ton prototype of the Jinping experiment, even in the worst working condition (the acrylic vessel is filled with water and no water outside the vessel), is only 2.1 MPa, which is still lower than 5 MPa. The maximum Mises stress on the stainless steel truss is 128.7 MPa, which is lower than 175 MPa. Meanwhile, the maximum truss deflection is 3.1 mm, which is approximately 1/635 of the span and smaller than the value specified in the code "Technical specification for space frame structures." The Buckle analysis of the stainless steel truss shows that the modal coefficient of the first buckling mode reaches 23.3, which is significantly higher than 4.2. The safety factor in consideration of the geometrical and material nonlinearities is equal to 6.9 that is significantly larger than 2.0. The maximum Mises stress on the stainless steel tank in normal working condition is equal to 5.5 MPa, which is much lower than the yield strength of the stainless steel. Consequently, the 1-ton prototype described in the present paper is safe and reliable from the mechanical point-of-view.

**Acknowledgement**

This study is financially supported in part by the National Natural Science Foundation of China (Nos. 51508424, 11235006, 11475093, and 11620101004) and the Tsinghua University Initiative Scientific Research Program of China (Nos. 20131089288, 20121088035, and 20151080432). We like to thank Jun Wang for making the final engineering design.